\documentclass[aps,twocolumn]{revtex4}
\usepackage[latin1]{inputenc}
\usepackage{graphicx}
\begin{document}

\title{Ab initio correlation approach to a ferric wheel-like molecular cluster}
\author{H. Nieber}
\email[e-mail: ]{harald.nieber@theochem.ruhr-uni-bochum.de}
\thanks{present address: Lehrstuhl für Theoretische Chemie, Ruhr-Universität Bochum, Universitätsstr. 150, 44801 Bochum}
\author{K. Doll}
\email[e-mail: ]{k.doll@tu-bs.de}
\author{G. Zwicknagl}
\affiliation{Institut f\"ur Mathematische Physik, TU Braunschweig, 
Mendelssohnstr. 3, 38106 Braunschweig, Germany}

\begin{abstract}

We present an \textit{ab initio} study
of electronic correlation effects
in a molecular cluster derived from
the hexanuclear ferric
wheel [LiFe$_6$(OCH$_3$)$_{12}$-(dbm)$_6$]PF$_6$.
The electronic and magnetic properties of this cluster
have been studied with all-electron Hartree-Fock,
full-potential density functional calculations
and multi-reference second-order perturbation theory.
For different levels of correlation,
a detailed study of the impact of the electronic correlation on the exchange parameter
was feasible.
As the main result, we found that
the influence of the bridge oxygen atoms
on the exchange parameter is less intense
than the influence of the apical ligand groups,
which is due to the geometry of the cluster.
With respect to the cluster model approach,
the experimental value of the exchange parameter
was affirmed.

\end{abstract}
\newpage
\maketitle
\section{Introduction}

In contemporary condensed-matter physics,
the role of molecular magnetism
is steadily growing and receives attention
from both experimental and theoretical
physicists and chemists \cite{sessoli1993,gatteschi1994,caneschi1999,pilawa1999}.
The most studied and therefore best known molecule in this
field is the Mn-12-acetate \cite{regnault2002},
but the class of the ferric wheels
(wheel-shaped iron rings) is becoming
an important subject of various studies\cite{schnack2004,waldmann1999,waldmann2001}.

In the past few years, molecular magnets
and some ferric wheels have been treated with \textit{ab initio} quantum chemical
methods like the Hartree-Fock approximation
and density functional approaches using
several functionals like the local density approximation
or hybrid functionals.
The usually observed results are that Hartree-Fock theory often
strongly underestimates physical properties like the exchange parameter \cite{nieber2005,towler1994,ricart1995,catti1995},
while the DFT methods overestimate those values \cite{nieber2005,postnikov2003,postnikov2004,iberio}.

This paper will deal with a molecular cluster derived
from the hexanuclear ferric
wheel [LiFe$_6$(OCH$_3$)$_{12}$-(dbm)$_6$]PF$_6$ \cite{nieber2005,Abbati1997}.
The previous analysis based on the full molecule \cite{nieber2005}
showed that one-determinantal Hartree-Fock
theory failed to reproduce the observed exchange
parameter $J$=-21 K \cite{Abbati1997}. Density functional calculations,
on the other hand, indicated that there is an enormous dependence of
the computed exchange parameter on the functional chosen.
This problem 
was already observed earlier for other systems, e.g.
\cite{MartinIllas1997,iberio,Iberio2004}.
A possible solution to this discrepancy is to
consider the electronic correlation, which is strongly influencing 
the exchange parameter, by wave function-based methods.
This should lead to a more controlled description of the magnetic behavior
 of the complex, as it was demonstrated for various systems
(e.g. \cite{casanovas1996,vanoosten1996,fink1,fink2,Graaf2001,Calzado2000,MoreiraKNiF,GraafKuprate,degraaf2004}).

\begin{figure}
\caption{Geometry of the molecular cluster. The model complex was derived from the full molecule
[LiFe$_6$(OCH$_3$)$_{12}$-(dbm)$_6$]PF$_6$ and was slightly modified to achieve $C_s(x)$-symmetry.}
\label{geometry}
\center\includegraphics[width=8cm,angle=0]{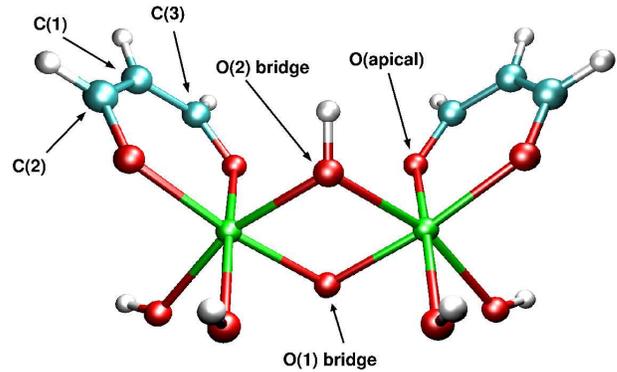}
\end{figure}

For antiferromagnetic systems like the ferric wheels,
the magnetic exchange can be qualitatively described
considering the electronic charge transfer
between the magnetic centers over a bridging atom or between
the magnetic centers and various ligand groups. Within a cluster approach,
the electronic charge transfer can be analyzed
within a multi-reference second-order perturbation theory scheme (MRPT2) \cite{celani2000}.
MRPT2 is very similar to the complete active space second-order perturbation theory (CASPT2) that
has been found to give very accurate results for the exchange parameter 
(see, e.g. \cite{degraaf2004,Graaf2001,iberio}).
The MRPT2 method is based on a reference ground state wave function
which can be chosen as a multiconfiguration self-consistent field (MCSCF) wave function.
This reference wave function is the initial point for treating the electronic correlation perturbationally.
In some cases, the occurrence of intruder states can cause convergence problems \cite{degraaf2004,hozoi2002},
and the level-shift technique proposed by Roos et al \cite{roos}
has to be applied.

In this paper,
we analyze the electronic properties of the simplified complex
displayed in Figure \ref{geometry}.
The model is derived from the molecule [LiFe$_6$(OCH$_3$)$_{12}$-(dbm)$_6$]PF$_6$
whose structure has been determined by Abbati et al \cite{Abbati1997}.
We 
first apply the Hartree-Fock and a hybrid functional (B3LYP) approach,
in order to verify the cluster model which is used to 
represent a fragment of the full molecule.
Then the effect of electronic correlations is studied more
detailed, with second order perturbation theory at the MRPT2
level.
The influence of the level shift on the MRPT2 results is determined in detail. 

\section{Method}\label{method}
\label{methodsection}
The geometry of the molecular cluster is
based on the measurements
of the primal ferric wheel
[LiFe$_6$(OCH$_3$)$_{12}$-(dbm)$_6$]PF$_6$ by Abbati et al \cite{Abbati1997}.
From these data, a complex consisting of two iron atoms
and some ligands was modeled (see Figure \ref{geometry}), as this
is the maximum what can be treated by MRPT2.
The iron atoms are six-fold coordinated and thus have the proper coordination
of the Fe ions like in the full molecule. Two point charges of +1 were
added at the position of the neighboring iron atoms in order to
restore the charge neutrality of the cluster.
The C$_6$H$_5$ rings
were replaced with hydrogen atoms (bonding length 0.93{\AA})
as well as the methyl groups at the bridge oxygen atoms (0.95{\AA}).
To achieve $C_s(x)$-symmetry which is necessary to keep the MRPT2
calculations tractable, the positions of the atoms were slightly
modified. The full geometry is given in Table \ref{coordinates}.

For a proper description of the physical properties
of the molecular cluster, calculations with the
codes CRYSTAL2003 \cite{crystal,dovesi} and MOLPRO2002 \cite{molpro}
were carried out.
Within the scope of the CRYSTAL calculations,
we employed the unrestricted Hartree-Fock (UHF) method
and the hybrid functional B3LYP
(a functional with admixtures, amongst others,
of functionals by Becke, Lee, Yang and Parr). Note that the CRYSTAL code
can treat systems of any periodicity, so that the molecular cluster 
was treated as a single molecule, i.e. not as a periodic system.
These calculations are of broken symmetry type
\cite{Noodleman1981,Caballol1997,Illas2000,Illas2004}, as 
the space symmetry is lowered. The state is not an eigenfunction
of $\mathbf{S^2}$, but only of $S_z$.

%\begin{widetext}
\begin{table}
\begin{center}
\caption{\label{coordinates}
Geometrical parameters of the molecular cluster, in {\AA}.
The mirror plane is the yz plane.}
\vspace{5mm}
\begin{tabular}{cccc}\vspace{0.3cm}
 &  x \hfill & y \hfill  & z \hfill  \\ \hline
Fe  &-1.568236  & 2.716264  & 0.000000 \\
O(1)  & 0.000000  & 1.952334  &-1.010690 \\
O(2)  & 0.000000  & 3.375149  & 1.066879 \\
H  & 0.000000  & 2.019278  &-1.966749 \\
H  & 0.000000  & 4.227152  & 1.505766 \\
O  &-1.660933  & 4.316006  &-1.146858 \\
O  &-2.906901  & 3.596739  & 1.146858 \\
C(3)  &-2.381312  & 5.371681  &-1.023420 \\
C(1) &-3.232182  & 5.601224  & 0.000000  \\
C(2) &-3.460321  & 4.748871  & 1.023420 \\
H &-4.021835  & 4.973378  & 1.723433 \\
H &-3.551594  & 6.475448  & 0.000000 \\
H &-2.293279  & 5.970805  &-1.723433  \\
O & -2.815456 & 1.465882  & -0.940540 \\
O & -1.442379 & 1.009753  & 1.120700 \\
H & -3.525746  & 1.728168 & -1.530737 \\
H & -1.465828  & 1.108581 & 2.107450 \\
+1 & -2.774296 & -0.177610 &  0.182920 \\ 
\end{tabular}
\end{center}
\end{table}
%\end{widetext}

For the needs of a molecular system,
we first chose the same basis set as used in ref. \cite{nieber2005}
for the full molecule, which was found to be reliable.
Henceforth this basis set
is referred as basis set A. In contrast to the code MOLPRO, the 
possibility of adding point charges is not implemented in the CRYSTAL code.
A point charge of +1 can however be achieved by
a H atom with a basis function with a very high exponent (100000 a.u.),
which does not allow charge transfer to the H atom
and thus acts like a point charge.
The idea of the CRYSTAL calculations was to verify that the results
did not significantly change when the geometry was modified from the
full ferric wheel with six iron atoms to the cluster
with two iron atoms. This was first done with basis set A, so that
an identical basis set was applied for the full molecule and the
fragment.  

With the code MOLPRO, calculations at the level of MCSCF
and MRPT2
were performed.
For these methods,
a modified basis set
(from now on labeled as basis set B) was used.
For iron, a [\textit{8s5p3d}] \cite{wachters} basis set
was chosen, where a $f$-exponent of 2.48 was added
which was optimized in a preceding calculation.
For oxygen, a [\textit{4s3p}] \cite{dunning} basis set
was chosen where a $d$-exponent of 0.8 was added.
The basis sets for carbon and hydrogen were chosen
accordingly to basis set A.
Thus,
the final basis set B was of the size
[\textit{8s5p3d1f}] (iron), [\textit{3s2p}] (carbon), [\textit{4s3p1d}]
(oxygen) and [\textit{2s}] (hydrogen).
The enlargement of the basis set
for iron and oxygen in basis set B
compared to basis set A is a necessary 
procedure to properly account for the needs
of \textit{post} Hartree-Fock calculations,
which in contrast to the calculations at the
UHF and B3LYP level include electronic excitations and thus
require a larger virtual orbital space.
To investigate the impact of the enlargement of the basis set on the results,
calculations were carried out with MOLPRO using either
basis set A or basis set B at the MCSCF level.
Despite the slightly different basis sets,
no significant changes in the results were observable,
and basis set B can be considered as an
extension of basis set A for wave function
based correlation calculations.
Subseqently all MRPT2 calculations were performed with the enlarged
basis set B.

\begin{figure}
\caption{Spin densities of the molecular cluster for the antiferromagnetic (AF) state at the UHF level (upper panel)
and the B3LYP level (lower panel).
Both graphs show the spin density in the plane
given by the planar arrangement of the six iron atoms of the primal
ferric wheel.
For all figures, the contour lines range from -0.0004 to 0.0005 in steps of 0.000035 electrons$/($a.u.$)^3$.
Full lines indicate positive spin density and dashed lines indicate negative spin density.}
\label{spindichte}
\center\includegraphics[width=7cm,angle=0]{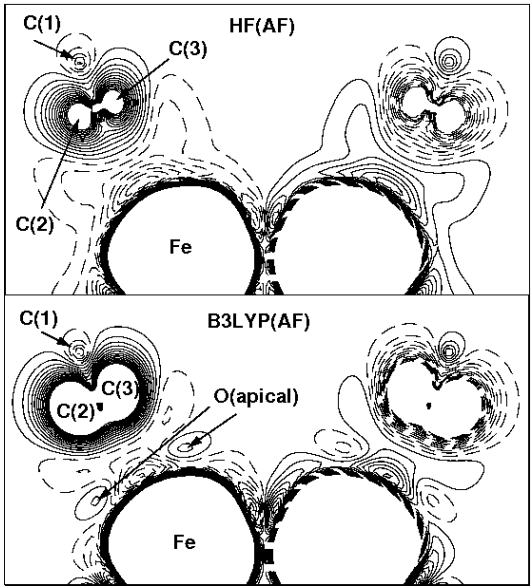}
\end{figure}

\begin{figure}
\caption{Magnetic exchange parameter of the molecular cluster for different values of the level shift.
The upper graph (squares) represents the uncorrected values,
the lower graph (triangles) the corrected values (correction performed according to Roos et al \cite{roos}).}
\label{levelshift}
\center\includegraphics[width=8cm]{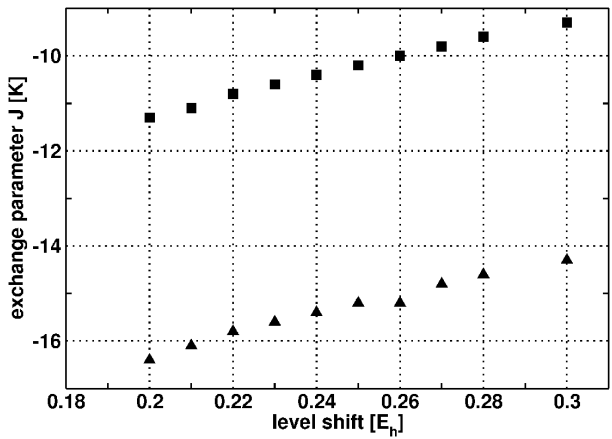}
\end{figure}

The properties
of the ferromagnetic (FM) state (all spins parallel, total spin 10 $\mu_B$)
and of the antiferromagnetic (AF) state (spins alternating up and down,
total spin 0) were computed with each method.
To obtain the net charge, a Mulliken population analysis was
performed at all levels of theory.

\section{Results}
\label{resultssection}

Table \ref{deltaenergy} summarizes the results for the total energies
of the ferromagnetic and the antiferromagnetic ground states,
their differences and the
magnetic exchange parameter $J$
for the molecular cluster.

Depending on the ansatz for the ground state wave functions
the magnetic coupling in the molecular cluster
is derived from fits to an Ising or Heisenberg
model.

The analysis in terms of a Heisenberg model

\begin{displaymath}
H_H = -J\cdot  \textbf{S}_1 \cdot \textbf{S}_2
\end{displaymath}

with

\begin{displaymath}
\textbf{S}_1\cdot \textbf{S}_2=\frac{(\textbf{S}_1+\textbf{S}_2)^2-\textbf{S}_1^2-\textbf{S}_2^2}{2}
\end{displaymath}

\noindent
is appropriate when the approximate ground state
is an eigenfunction of $(\textbf{S}_1+\textbf{S}_2)=\textbf{S}^2$.
This is the case for MCSCF and MRPT2.
The exchange parameter is obtained from the
difference between the ferromagnetic $(S=5)$
and the antiferromagnetic alignments which is given
by

\begin{eqnarray}
E_{H,tot}^{FM}-E_{H,tot}^{AF} & = &-J({\textbf{S}_1\cdot \textbf{S}_2}^{FM}-{\textbf{S}_1\cdot \textbf{S}_2}^{AF}) \nonumber \\
 & = & -15\cdot J \nonumber
\end{eqnarray}

\noindent
for quantum spins. The trial wave functions
used for UHF and B3LYP calculations
usually lack invariance under spin
rotation. Since they are constructed
as eigenfunctions of the spin projection $S_z$
the energy gain from magnetic correlations
is analyzed in terms of the Ising model:

\begin{displaymath}
H_I= -J\cdot  S_{1z} \cdot S_{2z}
\end{displaymath}

The corresponding energy difference
between ferromagnetic $(S_{1z}=S_{2z})$
and antiferromagnetic $(S_{1z}=-S_{2z})$
alignment is given by

\begin{displaymath}
E_{I,tot}^{FM}-E_{I,tot}^{AF}=-2\cdot J\cdot S_{1z}^2=-12.5\cdot J
\end{displaymath}

\noindent
which agrees with the Heisenberg value
in the classical limit $S_i\rightarrow\infty$.

The experimental value for the
exchange parameter of the primal ferric wheel was found to be
$J$=-21 K \cite{Abbati1997}.
Thus for all results in this dimension
the difference in the exchange parameter between the Ising and the Heisenberg model
is about a few Kelvin; with respect to this actuality
we consider the Ising model approach as a valid
description of the magnetic coupling in the cluster.

We first compare the results of
one-determinantal methods for the model cluster
and the full molecule. In a second step,
we focus on the influence of correlations.
In our previous studies of the
full molecule \cite{nieber2005}, we
determined an exchange parameter
of $J$=+7 K at the UHF level
and $J$=-31 K at the B3LYP level.

The UHF result for the full molecule
is perfectly reproduced by the cluster
yielding $J$=+7 K. A good agreement between cluster and
periodic system at the Hartree-Fock level was
already observed in earlier studies, e.g. NiO\cite{iberio},
KNiF$_3$ and K$_2$NiF$_4$
\cite{MoreiraKNiF} or Ca$_2$CuO$_3$ and Sr$_2$CuO$_3$ \cite{GraafKuprate}.
At the B3LYP level, the computed exchange parameter is $J$=-20 K. The
agreement with the value for the full molecule is
thus not as perfect as at the UHF level, but still reasonable
(a similar deviation was found, for example, when comparing B3LYP exchange 
couplings for K$_2$CuF$_4$ from cluster \cite{iberio1999} and periodic
systems \cite{Iberio2004}; it should be mentioned that there seem
to be fewer comparisons between cluster and periodic calculations
at the B3LYP level).
In addition, the Mulliken population analysis (Tables \ref{charge} and
\ref{fepop}) demonstrates that the charge is practically identical
at the various levels: B3LYP populations  for the cluster and
the full molecule \cite{nieber2005} are virtually identical, 
and similarly UHF populations agree very well. In addition, the
MCSCF charge agrees with the UHF charge.
As a whole, we feel that
the cluster model can be considered to be physically valid.

Concerning the individual charge, we note that
as a confirmation of the former findings \cite{nieber2005},
the net charge of iron is between 1.55 (B3LYP)
and 2.26 (MCSCF) and thus far away from the formal charge
of +3 of the primal ferric wheel \cite{Abbati1997}.
The charge is thus more delocalized at the B3LYP level, which
can also be seen in the spin-density plot in Figure \ref{spindichte}.
At the B3LYP level the local magnetic moment is distributed
over the cluster to a certain extent, while the level of delocalization is 
apparently smaller at the UHF level.

To serve as a starting point for the MRPT2 calculations,
a MCSCF calculation was performed.
The active space is made of the
iron $d$-orbitals. The obtained value for
the exchange parameter is $J$=-0.5 K.
Compared to UHF,
a MCSCF wave function is a better approximation
to the ground state than a single determinant,
which is the reason for the slight change of the exchange parameter
from UHF to MCSCF towards the experimental value.
A direct comparison between the calculations with the code CRYSTAL and MOLPRO
is only approximatively possible, and only in case of high spin. First,
the MOLPRO energy must be corrected for the fact that MOLPRO does not
take the interaction of point charges into account when computing the
total energy: when this is done, then the MCSCF energy with MOLPRO is
-3507.90171+1/(2*2.774296/.5291772) $E_h$=-3507.80634 $E_h$. The
remaining difference to the energy computed with CRYSTAL (-3507.80720 $E_h$)
is because the MOLPRO MCSCF wave function has identical orbitals
for up and down spin, whereas the CRYSTAL 
UHF wave function has not; and because
the codes use different screening parameters for the selection of the
integrals.

\begin{widetext}
\begin{table*}
\begin{center}
\caption{\label{deltaenergy}
Total energies ($E_h\equiv$ hartree), the differences in total energies of the ferromagnetic
(FM) and the antiferromagnetic (AF) state and exchange parameters \textit{J}.
At the MRPT2 level, all orbitals except the core orbitals were correlated
and a level shift of $0.3 E_h$ was applied.
The core orbitals include the iron \textit{1s-, 2sp-, 3sp}-, the oxygen \textit{1s}-
and the carbon \textit{1s}-orbitals.
For comparison, the results for the full molecule obtained from ref. \cite{nieber2005}
are given in the first section of this table.}
\vspace{5mm}
\begin{tabular}{ccccccc}\vspace{0.3cm}
basis set & molecule & method &  FM total energy \hfill & AF total energy 
\hfill & difference of & $J$ \hfill (K) \\
 & & & $(E_h)$ & $(E_h)$ & total energy \hfill $(mE_h)$
\\ \hline
A & full molecule \cite{nieber2005} & UHF & -13295.73287 & -13295.73125 & 
-1.62 & +7 \\
A & full molecule \cite{nieber2005}& B3LYP & -13334.50676 &  -13334.51409 
& 7.33 & -31 \\
 & & & & & \\ \hline
A & cluster & UHF & -3507.80720 & -3507.80693 & -0.27 & +7 \\
A & cluster & B3LYP & -3514.95130 & -3514.95209 & 0.79 & -20\\
 & & & & & &\\
A & cluster & MCSCF & -3507.90171 & -3507.90175 & +0.04 & -1 \\
B & cluster & MCSCF & -3508.31083 & -3508.31085 & +0.02 & -0.5 \\
 & & & & & &\\
B & cluster & MRPT2 & -3511.04152  & -3511.04209 & 0.57 & -14.4\\
\end{tabular}
\end{center}

\begin{center}
\caption{\label{charge} Charge at various sites, in $|e|$.
Note that the charge is virtually
identical for the ferromagnetic and the antiferromagnetic state.
For comparison, the results for the full molecule obtained from ref. \cite{nieber2005}
are given in the first section of this table.}
\vspace{5mm}
\begin{tabular}{ccccccc}\vspace{0.3cm}
basis set & molecule & method &  O (apical) \hfill     & O (bridge) \hfill & C(1) \hfill & C(2)/C(3) \hfill  \\ \hline
A & full molecule \cite{nieber2005} & UHF & -1.10 & -1.18 & -0.22 & 0.82 \\
A & full molecule \cite{nieber2005} & B3LYP & -0.86 & -0.88 & -0.11 & 0.62 \\
& & & & & \\ \hline
A & cluster & UHF & -1.03 & -1.13 & -0.17 & 0.77  \\
A & cluster & B3LYP & -0.80 & -0.89 & +0.01 & 0.60 \\
A & cluster & MCSCF & -1.03 & -1.14 & -0.17 & 0.77 \\
B & cluster & MCSCF & -1.15 & -1.22 & -0.13 & 0.85 \\
\end{tabular}
\end{center}
\end{table*}
\end{widetext}

Within the MCSCF scheme, the wave function is built from all Slater determinants
representing charge transfer configurations between the orbitals in the active space, and the orbitals are optimized.
The MRPT2 calculation is based on the MCSCF orbitals.
The reference configurations are made of all determinants in the active space (i. e. the iron $d$-orbitals),
and subsequently second-order perturbation theory is applied.
The level of correlation can additionally be varied by keeping different sets
of core orbitals frozen, as described later on in this section.

While performing the MRPT2 calculations,
intruder state problems
appeared which are due to a near
degeneracy of the ground state. One possibility
to remedy those intruder states is to increase
the active space, which is definitely not possible
for the considered molecular cluster,
or, following the proposal by Roos et al \cite{roos},
to implement a level shift to the MRPT2 calculations.
By means of this technique,
a level shift parameter is added to the
zeroth order Hamiltonian to avoid those intruder states.
Thus, the resulting exchange parameters are influenced by the level shift
(c.f. upper graph (squares) in Fig. \ref{levelshift}, corresponding to
equation 6 in \cite{roos}).
To approximatively correct for this effect,
a correction to the second order energies can be applied afterwards, as 
was suggested in \cite{roos}, equation 7.
The corresponding data are displayed in the lower graph (triangles) 
in Fig. \ref{levelshift}.
The quantitative dependence of the exchange parameter on the level
shift is shown in Table \ref{shiftlevel}.
For level shift values
smaller than 0.17 $E_h$ ($E_h\equiv$ hartree),
the MRPT2 results became unstable.
The level shift interval (for the corrected values of the exchange parameter)
from 0.20$E_h$ to 0.30$E_h$ leads to an
exchange parameter of $J$=15$\pm$2 K, which is a reasonably stable result.
For other systems, similar results were obtained, see e.g.
de Graaf et al \cite{degraaf2004} and Hozoi et al \cite{hozoi2002}.

%\twocolumngrid

%\begin{widetext}
\begin{table*}
\begin{center}
\caption{\label{fepop}
Mulliken charge of Fe, in $|e|$. Note that the charge is virtually
identical for the ferromagnetic and the antiferromagnetic state.
The $f$-population of the iron atoms is negligible.
The core orbitals include the iron \textit{1s-, 2sp-, 3sp}-, the oxygen \textit{1s}-
and the carbon \textit{1s}-orbitals.
For comparison, the results for the full molecule obtained from ref. \cite{nieber2005}
are given in the first section of this table.}
\vspace{5mm}
\begin{tabular}{ccccccc}\vspace{0.3cm}
basis set & molecule & method & net charge \hfill  &   \textit{s} \hfill & \textit{p} \hfill  & \textit{d} \\ \hline
A & full molecule \cite{nieber2005} & UHF & 2.16 & 6.28 & 12.28 & 5.29 \\
A & full molecule \cite{nieber2005} & B3LYP & 1.56 & 6.38 & 12.37 & 5.69 \\
& & & & & &\\ \hline
A & cluster & UHF & 2.13 & 6.29 & 12.30 & 5.28 \\
A & cluster & B3LYP & 1.55 & 6.39 & 12.40 & 5.66 \\
A & cluster & MCSCF & 2.14 & 6.29 & 12.30 & 5.27 \\
B & cluster & MCSCF & 2.26 & 6.37 & 12.07 & 5.31 \\
\end{tabular}
\end{center}
\end{table*}

\begin{table*}
\begin{center}
\caption{\label{shiftlevel}
Total energies ($E_h\equiv$ hartree), the differences in total energies for the ferromagnetic
(FM) and the antiferromagnetic (AF) state and exchange parameters \textit{J}
for different level shifts at the MRPT2 level for the molecular cluster. All orbitals except
the core orbitals were correlated.
The energies were corrected for the level shift.
The core orbitals include the iron \textit{1s-, 2sp-, 3sp}-, the oxygen \textit{1s}-
and the carbon \textit{1s}-orbitals.
Basis set B was used.}
\vspace{5mm}
\begin{tabular}{ccccc}\vspace{0.3cm}
 level shift &  FM total energy \hfill     $(E_h)$& AF total energy $(E_h)$\hfill & difference of total energy \hfill $(mE_h)$& $J$ \hfill (K) \\ \hline
0.20 & -3511.05498 & -3511.05563 & 0.65 & -16.4 \\
0.21 & -3511.05375 & -3511.05439 & 0.64 & -16.1 \\
0.22 & -3511.05251 & -3511.05314 & 0.63 & -15.9 \\
0.23 & -3511.05124 & -3511.05185 & 0.62 & -15.6 \\
0.24 & -3511.04993 & -3511.05054 & 0.61 & -15.4 \\
0.25 & -3511.04860 & -3511.04921 & 0.60 & -15.2 \\
0.26 & -3511.04724 & -3511.04784 & 0.59 & -15.0 \\
0.27 & -3511.04586 & -3511.04644 & 0.59 & -14.8 \\
0.28 & -3511.04444 & -3511.04502 & 0.58 & -14.6 \\
0.30 & -3511.04152 & -3511.04209 & 0.57 & -14.4 \\
\end{tabular}
\end{center}
\end{table*}

\begin{table*}
\begin{center}
\caption{\label{corrlevel}
Total energies ($E_h\equiv$ hartree), the differences in total energies for the ferromagnetic
(FM) and the antiferromagnetic (AF) state and exchange parameters \textit{J}
for different levels of correlation at the MRPT2 level for the molecular cluster. A level shift of $0.3 E_h$ was applied.
The energies were corrected for the level shift.
The core orbitals include the iron \textit{1s-, 2sp-, 3sp}-, the oxygen \textit{1s}-
and the carbon \textit{1s}-orbitals. Basis set B was used.}
\vspace{5mm}
\begin{tabular}{cccccc}\vspace{0.3cm}
 & level of correlation &  FM total energy \hfill     & AF total energy \hfill & difference of & $J$ \hfill (K) \\
 & & $(E_h)$ & $(E_h)$ & total energy \hfill $(mE_h)$ & \\
 \hline
(I) & iron \textit{d}-orbitals & -3508.43623 & -3508.43649 & 0.26 & -6.4  \\
(II) & (I)+\textit{2sp}-orbitals of O(1) & -3508.63134 & -3508.63154 & 0.19 & -4.9 \\
(III) & (I)+\textit{2sp}-orbitals of O(2) & -3508.64415 & -3508.64444 & 0.29 & -7.4 \\
(IV) & (I)+\textit{2sp}-orbitals of O(1),O(2) & -3508.84371 & -3508.84394 & 0.23 & -5.8 \\
(V) & (I)+all oxygen orbitals & -3510.58726 & -3510.58781 & 0.54 & -13.7 \\
(VI) & all orbitals except core & -3511.04152 & -3511.04209 & 0.57 & -14.4 \\
\end{tabular}
\end{center}
\end{table*}
%\end{widetext}

The results for the MRPT2 calculations are given in table \ref{corrlevel}.
Considering the magnetic iron $d$-orbitals
as the only orbitals to be correlated
led to an exchange parameter of \\ $J$=-6.4 K.
The included configurations are thus the charge transfer
configurations between the occupied iron
$d$-orbitals and the excitations
to the virtual orbitals (for the high-spin state,
only excitations to the virtual orbitals are possible).
Therefore, accounting for those charge transfer configurations
explains the change of the exchange parameter from the MCSCF level
to the MRPT2 level (-0.5K (MCSCF)$\rightarrow$ -6.4 K (MRPT2)).
Adding the $2sp$-orbitals of one of
the bridging oxygens to the orbitals to be correlated
gave rise to  an exchange parameter of $J$=-4.9 K (O(1))
and $J$=-7.4 K (O(2)).
Taking both these $2sp$-orbitals into
account caused only an exchange parameter of $J$=-5.8 K,
i. e. the effect of correlating both orbital groups
is approximatively additive.

Correlating all oxygen atoms results in a value of -13.7 K.
A further increase of the orbitals to be correlated
(up to a maximum where only the iron \textit{1s-, 2sp-, 3sp}-, the oxygen \textit{1s}-
and the carbon \textit{1s}-orbitals are kept frozen)
led to a exchange parameter of $J$=-14.4 K,
which is in the range of the experimental value. With respect
to the geometry of the cluster,
the influence of those apical ligand groups
to the exchange parameter is the dominant one, whereas
the oxygen
bridge atoms have only little impact on the exchange parameter.
As a conclusion, for this particular ferric wheel under
consideration the metal $\leftrightarrow$ ligand charge transfer configurations
dominate the metal $\leftrightarrow$ metal charge transfer 
configurations 
over the bridging oxygen atoms, which is basically the result of the
ordering of the magnetic orbitals according
to the Goodenough-Kanamori rules \cite{kahn}: essentially,
the Fe-O-Fe angle is nearly right-angled and thus the coupling is small.

In the experiments,
when comparing various ferric wheels, an approximatively linear
relationship between the Fe-O-Fe angle and the value of
the exchange coupling was observed \cite{waldmann2001} and
confirmed \cite{Pilawa2003}. This indicates that this angle
is crucial for the strength of the coupling (as long as the Fe-Fe distance is
approximatively constant, otherwise this distance may also have an impact). 
This is not in
contradiction to the findings here: essentially, the strength and
the nature of the coupling (ferro- or antiferromagnetic)
is strongly influenced by this angle, but still, to compute the
interaction properly, the ligands must be included in the correlation
treatment. This was demonstrated, for example, in \cite{Graaf2001}
(figure 2): when the correlation treatment is not sufficient,
the exchange couplings come out too small; but still, the dependence
on the angle is correct. Even more striking were earlier calculations
where the ligands were crucial to obtain
reasonable values for the exchange couplings, e.g. for KNiF$_3$ and 
K$_2$NiF$_4$
\cite{MoreiraKNiF} or NiO \cite{deGraaf1997JCP}. 

\section{Conclusion}
\label{summarysection}

Wave function-based correlation methods were applied
to a molecular cluster derived from the hexanuclear
ferric wheel [LiFe$_6$(OCH$_3$)$_{12}$-(dbm)$_6$]PF$_6$ \cite{Abbati1997}.
The validity of the molecular cluster containing two iron atoms was tested
by means of a one-determinantal approach
with respect to the formerly calculated results for the full molecule
\cite{nieber2005} 
for the exchange parameter of the primal ferric wheel.
In addition, at the UHF and B3LYP level,
the spin densities and the electronic population,
at the MCSCF level the
electronic population were calculated.
The population analysis supported the validity
of the cluster model approach.

The best result for the exchange coupling parameter $J$
was obtained at the MRPT2 level ($J$=15$\pm$2 K). 
The influence of intruder state problems
on the exchange parameter was explicitly investigated, and
applying the level shift technique \cite{roos} was found
to lead to stable results. MRPT2 thus gives
a more controlled approach to the  importance of electronic correlations
for exchange couplings, whereas the density functional results depend
strongly on the functional chosen.
Also, the impact of certain atom groups on the exchange parameter
was determined at the MRPT2 level.
Correlation of the electrons of the
bridging oxygen atoms was of minor importance for the 
coupling strength. A strong enhancement of the computed exchange coupling was
however observed by additionally 
correlating the electrons of the apical oxygen atoms.

\section{Acknowledgments}

Most of the calculations were performed at the compute-server \textit{cfgauss}
(Compaq ES 45) of the data processing center of the TU Braunschweig.
The geometry plot of the molecular cluster was performed with VMD \cite{vmd}.


\begin{references}
\bibitem{sessoli1993} R. Sessoli, D. Gatteschi, A. Caneschi and M. A. Novak, Nature {\bf365}, 141 (1993)
\bibitem{gatteschi1994} D. Gatteschi, A. Caneschi, L. Pardi and R. Sessoli, Science {\bf 265}, 1054 (1994)
\bibitem{caneschi1999} A. Caneschi, D. Gatteschi, C. Sangregorio,
R. Sessoli, L. Sorace, A. Cornia, M. A. Novak, C. Paulsen and W. Wernsdorfer,
J. Mag. Mag. Mat., {\bf 200} 182 (1999)
\bibitem{pilawa1999} B. Pilawa, Ann. Phys. (Leipzig), {\bf 8}, 191 (1999)
\bibitem{regnault2002} N. Regnault, T. Jolicoeur, R. Sessoli, D. Gatteschi and M. Verdaguer,
Phys. Rev. B {\bf66}, 054409 (2002)
\bibitem{schnack2004} J. Schnack, Lect. Notes Phys. {\bf645}, 155 (2004)
\bibitem{waldmann1999} O. Waldmann, J. Sch\"ulein, R. Koch, P. M\"uller,
I. Bernt, R. W. Saalfrank, H. P. Andres, H. U. G\"udel and P. Allensbach,
Inorg. Chem. {\bf 38}, 5879 (1999)
\bibitem{waldmann2001} O. Waldmann, R. Koch, S. Schromm, J. Schülein, P. Müller, I. Bernt, R. W. Saalfrank, F. Hampel and E. Balthes, Inorg. Chem. 
\textbf{40}, 2986 (2001)
\bibitem{nieber2005} H. Nieber, K. Doll and G. Zwicknagl, Eur. Phys. J. B {\bf 44}, 209 (2005)
\bibitem{towler1994}
M.D. Towler, N.L. Allan, N.M. Harrison, V.R. Saunders, W.C. Mackrodt and
E. Apr\`a, Phys. Rev. B {\bf 50}, 5041 (1994)
\bibitem{ricart1995}J. M. Ricart, R. Dovesi, C. Roetti and V. R. Saunders,
Phys. Rev. B {\bf 52}, 2381 (1995)
\bibitem{catti1995} M. Catti, R. Valerio and R. Dovesi, Phys. Rev. B {\bf 51},
7441 (1995)
\bibitem{postnikov2003} A. V. Postnikov, J. Kortus and S. Bl\"ugel, Mol.
Phys. Rep. {\bf 38}, 56 (2003)
\bibitem{postnikov2004} A. V. Postnikov, S. G. Chiuzb\u{a}ian, M. Neumann and
S. Bl\"ugel, J. Phys. Chem. Solids {\bf 65}, 813 (2004)
\bibitem{iberio}I. de P. R. Moreira, F. Illas and R. L. Martin, Phys. Rev. B
{\bf 65}, 155102 (2002)
\bibitem{Abbati1997} G. L. Abbati, A. Cornia, A. C. Fabretti, W. Malavasi, L. Schenetti, A. Caneschi and D. Gatteschi, Inorg. Chem. {\bf 36}, 6443 (1997)
\bibitem{MartinIllas1997} R. L. Martin and F. Illas, Phys. Rev. Lett.
{\bf 79}, 1539 (1997)
\bibitem{Iberio2004} I. de P. R. Moreira and R. Dovesi,
Int. J. Quant. Chem. {\bf 99}, 805 (2004)
\bibitem{casanovas1996} J. Casanovas, J. Rubio and F. Illas, Phys. Rev. B {\bf 53}, 945 (1996)
\bibitem{vanoosten1996} A. B. van Oosten, R. Broer and W. C. Nieuwpoort, Chem. Phys. Lett. {\bf 257}, 207 (1996)
\bibitem{fink1} K. Fink, R. Fink and V. Staemmler, Inorg. Chem. {\bf 33}, 6219 (1994)
\bibitem{fink2} K. Fink, C. Wang and V. Staemmler, Inorg. Chem. {\bf 38}, 3847 (1999)
\bibitem{degraaf2004} C. de Graaf, L. Hozoi and R. Broer, J. Chem. Phys. {\bf 120}, 961 (2004)
\bibitem{Graaf2001} C. de Graaf, C. Sousa, I. de P. R. Moreira and
F. Illas, J. Phys. Chem. A {\bf 105}, 11371 (2001)
\bibitem{Calzado2000} C. J. Calzado, J. F. Sanz and J. P. Malrieu,
J. Chem. Phys. {\bf 112}, 5158 (2000)
\bibitem{MoreiraKNiF} I. de P. R. Moreira and F. Illas, Phys. Rev. B
{\bf 55}, 4129 (1997)
\bibitem{GraafKuprate} C. de Graaf and F. Illas, Phys. Rev. B
{\bf 63}, 014404 (2000)
\bibitem{celani2000}P. Celani and H.-J. Werner, J. Chem. Phys. {\bf 112}, 5546 (2000)
\bibitem{hozoi2002} L. Hozoi, A. H. de Vries, A. B. van Oosten, R. Broer, J. Cabrero and C. de Graaf, Phys. Rev. Lett. \textbf{89}, 076407 (2002)
\bibitem{roos} B. O. Roos and K. Andersson, Chem. Phys. Lett. {\bf 245}, 215 (1995)
\bibitem{crystal} V. R. Saunders, R. Dovesi, C. Roetti, R. Orlando,
C. M. Zicovich-Wilson, N. M. Harrison, K. Doll, B. Civalleri, I. J. Bush,
P. D'Arco and M. Llunell, \textbf{CRYSTAL2003} User's Manual 2003
\bibitem{dovesi}R. Dovesi, R. Orlando, C. Roetti, C. Pisani and V. R. Saunders,
phys. stat. sol. (b) \textbf{217}, 63 (2000)
\bibitem{molpro}
H.-J. Werner, P. J. Knowles, R. Lindh, M. Schütz, P. Celani, T. Korona, F. R. Manby, G. Rauhut, R. D. Amos, A. Bernhardsson,
A. Berning, D. L. Cooper, M. J. O. Deegan, A. J. Dobbyn, F. Eckert, C. Hampel, G. Hetzer, A. W. Lloyd, S. J. {McNicholas}, W. Meyer,
M. E. Mura, A. Nicklass, P. Palmieri, R. Pitzer, U. Schumann, H. Stoll, A. J. Stone, R. Tarroni and T. Thorsteinsson,
\textbf{MOLPRO} (version 2002.6), a package of ab initio programs (2003)
\bibitem{Noodleman1981} L. Noodleman, J. Chem. Phys. {\bf 74}, 5737 (1981)
\bibitem{Caballol1997} R. Caballol, O. Castell, F. Illas, I. de P. R.
Moreira, J. P. Malrieu, J. Phys. Chem. A {\bf 101}, 7860 (1997)
\bibitem{Illas2000} F. Illas,  I. de P. R. Moreira, C. de Graaf and
V. Barone, Theor. Chem. Acc. {\bf 104}, 265 (2000)
\bibitem{Illas2004} F. Illas,  I. de P. R. Moreira, J. M. Bofill and 
M. Filatov, Phys. Rev. B {\bf 70}, 132414 (2004)
\bibitem{wachters} A. J. H. Wachters, J. Chem. Phys. \textbf{52}, 1033 (1970)
\bibitem{dunning} T. H. Dunning, Jr. J. Chem. Phys. \textbf{90}, 1007 (1989)
\bibitem{iberio1999} I. de P. R. Moreira and F. Illas, Phys. 
Rev. B {\bf 60}, 5179 (1999)
\bibitem{kahn} O. Kahn, \textbf{Molecular Magnetism}, VCH Publishers Inc., New York (1993)
\bibitem{Pilawa2003} B. Pilawa, I. Keilhauer, G. Fischer, S. Knorr, J. Rahmer
and A. Grupp, Eur. Phys. J. B {\bf 33}, 321 (2003)
\bibitem{deGraaf1997JCP} C. de Graaf, F. Illas, R. Broer and W. C. Nieuwpoort,
J. Chem. Phys. {\bf 106}, 3287 (1997)
\bibitem{vmd} W. Humphrey, A. Dalke and K. Schulten, J. Molec. Graphics {\bf 14}, 33 (1996) 


\end{references}
\end{document}